\documentclass[12pt,a4paper]{article}

\usepackage{epsfig}
\usepackage{amsmath}

\def\nostrocostrutto#1\over#2{\mathrel{\mathop{\kern 0pt \rlap 
  {\raise.2ex\hbox{$#1$}}}
  \lower.9ex\hbox{\kern-.190em $#2$}}}
\def\lsim{\nostrocostrutto < \over \sim}   

\catcode`@=11
\newcount\@tempcntc
\def\@citex[#1]#2{\if@filesw\immediate\write\@auxout{\string\citation{#2}}\fi
  \@tempcnta\z@\@tempcntb\m@ne\def\@citea{}\@cite{\@for\@citeb:=#2\do
    {\@ifundefined
       {b@\@citeb}{\@citeo\@tempcntb\m@ne\@citea\def\@citea{,}{\bf ?}\@warning
       {Citation `\@citeb' on page \thepage \space undefined}}%
    {\setbox\z@\hbox{\global\@tempcntc0\csname b@\@citeb\endcsname\relax}%
     \ifnum\@tempcntc=\z@ \@citeo\@tempcntb\m@ne
       \@citea\def\@citea{,}\hbox{\csname b@\@citeb\endcsname}%
     \else
      \advance\@tempcntb\@ne
      \ifnum\@tempcntb=\@tempcntc
      \else\advance\@tempcntb\m@ne\@citeo
      \@tempcnta\@tempcntc\@tempcntb\@tempcntc\fi\fi}}\@citeo}{#1}}
\def\@citeo{\ifnum\@tempcnta>\@tempcntb\else\@citea\def\@citea{,}%
  \ifnum\@tempcnta=\@tempcntb\the\@tempcnta\else
   {\advance\@tempcnta\@ne\ifnum\@tempcnta=\@tempcntb \else \def\@citea{--}\fi
    \advance\@tempcnta\m@ne\the\@tempcnta\@citea\the\@tempcntb}\fi\fi}
\catcode`@=12

\begin{document}

\setcounter{page}{0}
\thispagestyle{empty}

\vspace*{-1cm}
\hfill \parbox{3.5cm}{BUTP-03/10 \\ MPI-PhT/2003-16\\
April 15, 2003}
\vfill

\begin{center}
  {\large {\bf
Scalar  mesons and glueball in $B$-decays and gluon jets\footnote{Work
      supported in part by the Schweizerischer Nationalfonds.}}  }
\vfill
{\bf
    Peter Minkowski } \\
    University of Bern \\
    CH - 3012 Bern, Switzerland
   \vspace*{0.3cm} \\  
   and \vspace*{0.3cm} \\
{\bf
    Wolfgang Ochs } \\
    Max-Planck-Institut f\"ur Physik \\
    Werner-Heisenberg-Institut \\
    D - 80805 Munich, Germany\\  


\vfill
\begin{abstract}
\noindent
We discuss the recent observation of $f_0(980)$
in charmless $B$-decays and in gluon jets which hints toward
a gluonic coupling of this meson similar to $\eta'$. 
Further predictions on $B$-decays into scalar particles are presented.
Charmless $B$ decays also show 
a broad $K\overline K$ 
(and possibly $\pi\pi$) 
$S$-wave enhancement
which we relate to the $0^{++}$ glueball. 
These gluonic mesons represent a sizable fraction of the 
theoretically derived decay rate for
$b\to sg$.
\end{abstract}
\end{center}  

\vfill



\newpage
\section{Introduction}
There is still no general consensus about the lightest 
scalar $q\overline q$ nonet, neither about it's members nor about
the mixing between strange and nonstrange components, also the existence
and mixing properties of the $J^{PC}=0^{++}$ glueball are in doubt. 
A central role for the nonet is played by $f_0(980)$ which
has been considered not only as standard $q\overline q$ meson 
but also as $K\overline K$-molecule or as 4-quark state. 

The spectroscopic data often have not been precise enough to arrive at unique
conclusions. New results on $D$ and $B$ decays of high statistics 
are now providing additional information. In this paper we discuss
some recent experimental results and their implications on these
problems:\\
\vspace*{-0.5cm}
\begin{description}
\item 1. The observation by the BELLE collaboration of charmless decays $B\to Khh$
with $h=\pi,K$ \cite{belle,belle2} which show a significant signal 
of $f_0(980)$; this has been observed recently also by the BaBar
collaboration \cite{babar1}.\\
\vspace*{-0.7cm}
\item 2. In the same channel BELLE has also observed a 
broad enhancement in $K\overline K$ mass spectrum in the range 1000-1700
MeV with  spin $J=0$ and a smaller effect in $\pi\pi$ around 1000 MeV.
Preliminary results from BaBar 
\cite{babar2} confirm this effect 
but there is no quantitative analysis yet.\\
\vspace*{-0.7cm}
\item 3.  A significant signal of $f_0(980)$, larger than expected, 
has also been observed in a first analysis of
the leading system in gluon jets obtained by DELPHI at LEP \cite{bm}. 
\end{description}
The interest in charmless  $B$-decays with strangeness has been stimulated
through the observation by
CLEO \cite{cleo0,cleo1} of large inclusive and exclusive 
decay rates $B\to \eta'X$ and $B\to \eta'K$, which have been
confirmed by more recent measurements \cite{cleo2,belle1,babar}. These
processes have been related to the decay  $b\to sg$ of the $b$-quark 
which could be a source of mesons with large gluon affinity
\cite{soni,fritzsch,hou,dgr}. In consequence, besides $\eta'$ 
also other gluonic states, in particular also scalar mesons or glueballs  
could be produced in a similar way. 

The total rate $b\to sg$  has been calculated  perturbatively
in leading \cite{ciuchini} and next-to-leading order \cite{greub}
\begin{equation}
\text{Br} (b\to sg) = 
\begin{cases}
(2-5)\times 10^{-3} & \text{in LO  (for $\mu=m_b\ldots m_b/2$)}\\
(5\pm 1)\times 10^{-3} &  \text{in NLO}
\end{cases}
\label{btosg}
\end{equation} 
The energetic massless gluon in this process could turn
entirely into gluonic mesons by a nonperturbative transition
after colour neutralization by a second gluon.
Alternatively, colour neutralization through $q\overline q$ pairs is
possible
as well. This is to be distinguished from the short distance process $b\to   
s\overline q q$ with virtual intermediate gluon
which has to be added to the
CKM-suppressed decays $b\to q_1\overline q_2 q_2$. These quark processes
with $s$
have been calculated and amount to branching fractions of 
$\sim 2\times 10^{-3}$ each
\cite{altarelli,nierste,greub}.
The question then arises which 
hadronic final states correspond to the decay  $b\to sg$.

Here we discuss how the above
new results and further measurements can clarify the low
mass spectroscopy of scalar particles and their contribution to the gluonic
$B$ decays.
In a previous study \cite{mo} we have performed 
a detailed phenomenological analysis of production
and decay of low mass scalar mesons, which led us to identify 
the scalar nonet with the states
$a_0(980),\ f_0(980),\ K^*_0(1430)$ and $f_0(1500)$
 with large flavour
mixing, just as in the pseudoscalar nonet. The near flavour
singlet states are the parity partners $\eta'$ and
$f_0(980)$  whereas near flavour octet states are $\eta$ and 
$f_0(1500)$. This scalar nonet fulfills the Gell Mann-Okubo mass formula
and is also consistent with a general QCD potential model.
The left over states $f_0(400-1200)$ (also called
$\sigma(600)$) and $f_0(1370)$ seen in $\pi\pi$ and other channels 
have been interpreted as signals from a single broad object
centered around 1 GeV with a large width of   
500-1000 MeV which we take as the $0^{++}$ glueball.
In this paper we discuss how this scheme compares with the new
data and how further measurement could clarify the structure of
the scalar sector. 

There are alternative schemes for low mass 
$q\overline q$ and glueball spectroscopy
which include: light $q\overline q$ nonet like ours, 
except for $a_0(980)$ but no glueball \cite{klempt}; 
QCD sum rule analysis \cite{narison}
with  $f_0(980)$ and broad $\sigma$ around 1000 MeV, 
both mixed in equal parts from glueball and light quark scalar;
a broad glueball in the range
1000-1600 MeV from overlapping $f_0$ states in $K$ matrix analyses 
(recent results \cite{anis}) but with $f_0(980)$ near flavour octet. 
$B$ decays may clarify these alternatives.

\section{Charmless $B$ decays with $K$ and $K^*(890)$}
{\it Two-body decays into pseudoscalar and vector mesons PP and PV}\\
We begin by reconsidering the decays $B\to K\eta',\ K^*\eta'$ 
together with other final states related 
by $U(3)$ symmetry. Subsequently we wish to extend these considerations 
to the inclusion of scalar particles.
The large branching fraction $B\to K\eta'$
confirms the special role of $\eta'$ in these
decays and it has been related \cite{soni,fritzsch,hou,dgr} to the gluon
affinity of $\eta'$, especially 
through the QCD axial anomaly which affects only the flavour singlet
component. However, it
appears difficult to explain the $K\eta'$ rate entirely by quark final
states and the QCD anomaly within a perturbative framework \cite{acgk},
a factor 2 remains unexplained. An improvement is possible by inclusion of
radiative corrections \cite{bn} but with considerable uncertainties.

Alternatively, one may introduce a phenomenological flavour singlet
amplitude which allows also for non-perturbative effects \cite{dgr}.
This amplitude is added to the dominant 
penguin amplitudes, 
 the small 
 tree amplitudes and electroweak penguins. Different decays are related by
flavour $U(3)$ symmetry.
 Recent applications \cite{cr} of this scheme to 2-body $B$ decays  
with strange and nonstrange pseudoscalar and vector  
particles yield a good overall
agreement with the data in terms of a few phenomenological input amplitudes. 

Here we discuss
 first the 2-body $B$ decays with $K$ and $K^*$ in this way \cite{cr}
to understand the pattern of the observed rates and then extend the analysis
to the scalar sector. 
For this purpose we restrict ourselves to a simple approximation and at
short distances we keep only the dominant QCD 
penguin amplitudes $T_q$ for $b\to u\overline u s,
d\overline d s, s\overline s s$ with $T_u= T_d=T_s$. 
The hadronic penguin amplitude $p_{AB}$ for the 2-body $B$ decay
into  particles from $U(3)$ multiplets $A$ and $B$ 
are then proportional to the superposition of short distance amplitudes
$T_q$ corresponding to the quark composition of the final hadrons; in
addition, there is the contribution from the flavour singlet amplitude
which we write as $\gamma_{AB} p_{AB}$.
The quark mixing in the pseudoscalar sector is taken as 
$\eta=(u\overline u + d\overline d - s\overline s)/\sqrt{3}$ and
$\eta'=(u\overline u + d\overline d + 2 s\overline s)/\sqrt{6}$,
see Table \ref{tab:pseudo}.

If the two particles belong to two different multiplets then $p_{AB}$
is the  amplitude for the $A$ particle carrying the $s$ quark from 
 $b\to sg$ decay 
and the $B$ particle carrying the spectator quark $q_s$.
 The amplitude for both particles interchanged, i.e. $s\to B,\ q_s\to A$,
is written as $\beta_{AB} p_{AB}$. 
In case of $B$-decays into mesons with $s\overline s$ component
($\eta,\eta',\phi\ldots$) both amplitudes contribute and
interfere
with full amplitude $p_{AB}(1+(-1)^L \beta_{AB})$ where the second term with
$\beta'=(-1)^L\beta$ refers to the 2-particle state with reflected momenta
($\vec p \to -\vec p$) for orbital angular momentum $L$
\cite{lipkin}. In particular, for 
$B\to VP$ decays there is a relative $(-)$ sign for the interchange amplitude.

Concerning our leading penguin approximation 
we note the following:
the decay rates at the quark level including penguin and
tree amplitudes have been calculated \cite{nierste,greub} and the quark
decays of interest to us 
are found with relative fractions \cite{nierste} 22\% ($b\to u\overline u
s$), 18\%
($b\to d\overline d s$) and 15\% ($ b\to  s\overline s s$)
 of all charmless $B$ decays. So 
the nonleading weak decay amplitudes which we neglect here 
modify the leading result from penguins by about $\pm20\%$. 
Effects of that size are also found in the phenomenologically determined
hadronic non-penguin amplitudes \cite{cr}.

\begin{table}[p]
\caption{Branching ratios for $B^+$ and $B^0$ decays into pseudoscalar (P) and
vector (V) particles (col. 4-6) in terms of amplitudes $T_q$ (col. 2) 
for  decays
$b\to s \overline q q$,
$\gamma,\gamma'$ and $\beta$ for gluonic and interchange processes, 
 col. 3: $p_{AB}$ set to 1, 
col. 5: $\alpha=0.67,\
\gamma=0.53$ (always $\beta=-\beta'=1$, $\gamma=\gamma'$),  see also text.}
\vspace*{0.1cm}
$
\begin{array}{llcllc}
\hline\hline
B\to PP & \text{amplitudes} & T_q=1 & \gamma=0 & 
\alpha,\gamma &
              \text{Br}_{\text{exp}} [10^{-6}]\\
\hline
K^0\pi^+ & T_d & 1 & \text{input} &  \text{input} & 17.3^{+2.7}_{-2.4} 
\\
K^+\pi^0 & \frac{1}{\sqrt{2}} T_u & \frac{1}{\sqrt{2}} & 8.7 &  8.7 & 
     12.1\pm 1.6\\
K^+\eta & \frac{1}{\sqrt{3}} (T_u-T_s+\gamma T_d) & \frac{\gamma}{\sqrt{3}} 
     & 0.0 &  1.6 & <6.9
\\
K^+\eta' & \frac{1}{\sqrt{6}} (T_u+2T_s+4\gamma T_d) &
\frac{3+4\gamma}{\sqrt{6}}      & 26.0 &  \text{input}\ \gamma & 75\pm 7
\\
\hline
K^+\pi^- & T_u & 1 & 15.9 &  15.9 & 17.4\pm 1.5
\\
K^0\pi^0 & \frac{1}{\sqrt{2}} T_d & \frac{1}{\sqrt{2}} & 8.0 & 8.0 &
   10.7^{+2.7}_{-2.5} 
\\
K^0\eta & \frac{1}{\sqrt{3}} (T_d-T_s+\gamma T_d) & \frac{\gamma}{\sqrt{3}} 
     & 0.0 &  1.5 & <9.3
\\
K^0\eta' & \frac{1}{\sqrt{6}} (T_d+2T_s+4\gamma T_d) &
\frac{3+4\gamma}{\sqrt{6}}      & 23.9 &  69.4 &   58^{+14}_{-13}
\\
\hline\hline
B\to VP & &\alpha=1 &&&\\
\hline
K^{*0}\pi^+ & \alpha T_d &  1 & 17.3 & 7.9 & 19^{+6}_{-8} 
\\
K^{*+}\pi^0 &  \frac{\alpha}{\sqrt{2}} T_u & \frac{1}{\sqrt{2}} & 8.7 &
   3.9 & <31 
\\
K^{*+}\eta & \frac{\alpha}{\sqrt{3}} (T_u-\beta' T_s+\gamma' T_d) &
  \frac{2+\gamma'}{\sqrt{3}}
     & 23.1 &  36.9 &  26^{+10}_{-9}
\\
K^{*+}\eta' & \frac{\alpha}{\sqrt{6}} (T_u+2\beta' T_s+4\gamma' T_d) &
\frac{-1+4\gamma'}{\sqrt{6}}  & 2.9 &  3.6 & <35
\\
\rho^+ K^0 & \alpha\beta T_d & 1 & 17.3 & 7.9 & <48 \\
\rho^0 K^+ &  \frac{\alpha\beta}{\sqrt{2}}T_u & \frac{1}{\sqrt{2}}&
     8.7 & 4.0 & <12 \\
\omega K^+&  \frac{\alpha\beta}{\sqrt{2}}T_u & \frac{1}{\sqrt{2}}&
      8.7 & 4.0 & <4 \\
\phi K^+ & -\alpha T_s & 1 & 17.3 & \text{input}\ \alpha & 7.9^{+2.0}_{-1.8}\\
\hline
K^{*+}\pi^- & \alpha T_u & 1 & 15.9 &  7.3 & <72 
\\
K^{*0}\pi^0 & \frac{\alpha}{\sqrt{2}} T_d & \frac{1}{\sqrt{2}} & 8.0 & 3.6 &
   <3.6
\\
K^{*0}\eta & \frac{\alpha}{\sqrt{3}} (T_d-\beta' T_s+\gamma' T_d) & 
    \frac{2+\gamma'}{\sqrt{3}}   & 21.3 &  15.7 &  14^{+6}_{-5}
\\
K^{*0}\eta' & \frac{\alpha}{\sqrt{6}} (T_d+2\beta' T_s+4\gamma' T_d) &
\frac{-1+4\gamma'}{\sqrt{6}}      & 2.6 &  1.5&  <24
\\
\rho^- K^+& \alpha\beta T_u & 1 & 15.9 & 7.3  & <32 \\
\rho^0 K^0&  \frac{\alpha\beta}{\sqrt{2}}T_d & \frac{1}{\sqrt{2}}&
      8.0 & 3.6 & <3.9 \\
\omega K^0&  \frac{\alpha\beta}{\sqrt{2}}T_d & \frac{1}{\sqrt{2}}&
      8.0 & 3.6 & <13 \\
\phi K^0 & -\alpha T_s & 1 & 15.9 & 7.3 & 7.6\pm 1.4\\
\hline\hline
\end{array}
$
\label{tab:pseudo}
\end{table}

The results in our leading approximation are given in Table
\ref{tab:pseudo}. There are two adjustable (penguin) amplitudes, chosen real, 
for both multiplets
 $p_{PP}$ and $p_{VP}$ ($\alpha=  p_{VP}/p_{PP}$),
the flavour singlet amplitudes $\gamma p_{PP}$ and $\gamma' p_{VP}$
($\gamma\equiv \gamma_{PP},\ \gamma'\equiv\gamma_{VP}$)
 and the interchange amplitude in case of $VP$ decay $\beta p_{VP}$.
First we try  
the simplest approximation 
with equal strength for both
multiplets ($\alpha=1$) and equal recombination $\beta=1$, 
also $\gamma=0$. 
Then all branching
ratios  are given in terms of one overall normalization
parameter ($p_{PP}$). The corresponding predictions are given in column 3 
(in units of $p_{PP}$ and $p_{VP}$) and 4
of Table \ref{tab:pseudo}. The predictions for $B^0$ are obtained after
multiplying $|p_{AB}|^2$ by the ratio $\tau_{B^0}/\tau_{B^+}=0.921$. We compare
with experimental data compiled by the PDG \cite{pdg} which includes 
$\eta$ and $\eta'$ decays \cite{cleo2,belle1,babar}.

One can see that for $PP$ decays the overall
pattern is reproduced, except for $K\eta'$ which is observed 
significantly too large by a
factor $\sim 3$. In this scheme this conclusion is derived from 
flavour symmetry,
the neglect of nonleading short distance terms was only about
20\%. Agreement with data can be obtained by adding
the flavour singlet amplitude with $\gamma=0.53$ which predicts also effects for
 $\eta$. 
Too large predictions are found in this approximation 
for the $VP$ decays $\phi K$, $K^{*0}\pi^0$, $\omega K^+$ and $\rho K^0$. 
A 
solution is possible by choosing a different normalization $\alpha=0.67$ 
for $VP$ decays keeping $\gamma'=\gamma$.

With this choice the model can also reproduce the decay pattern of 
 $B\to K^*(890)+(\pi,\eta,\eta')$ with reversed abundances of $\eta$ and
$\eta'$. This is a consequence of the different sign of $\beta'$ in the
PP and VP amplitudes \cite{lipkin,cr}, a feature also 
present in other analyses \cite{acgk,bn} for the same reason.
 One can estimate the gluonic part of the $\eta' K^+$ production 
alone (without interference)
 from contributions $\sim |\gamma|^2$ 
 to $\eta K^+$ and $\eta' K^+$ rates and obtains
\begin{equation}
{\rm Br}(B^+\to \eta' K^+)|_{\text{gluonic}} = 
(8/3)\ |\gamma\ p_{PP}|^2 
\sim (15\ldots 35)\times 10^{-6}
\label{etaprgluon}
\end{equation}
where the smaller
number refers to real $\gamma=0.53$ 
and the second one to arbitrary gamma with $|\gamma|=0.88$
 (see also \cite{cr}), 
$Re \gamma<0$ would be in conflict with the $K^+\eta$ rate.

At this level of approximation, accurate to $\sim20\%$, 
there are no major discrepencies encountered.
We conclude that the main effects are
reproduced by 3 parameters, the two penguin amplitudes $p_{PP}$ and $p_{VP}$
and the gluonic amplitude with $\gamma$, 
furthermore we have chosen $\beta=1$ and $\gamma'=\gamma\equiv \gamma_P$.

{\it Decay  $B\to f_0(980)K$ and expectations for scalar particles}\\
A remarkably strong signal is observed for the scalar meson $f_0(980)$
by the BELLE Collaboration \cite{belle} in the decay $B^+\to K^+\pi^+\pi^-$ 
where almost one half of the total rate above background falls into this
sub-channel with
\begin{equation}
{\rm Br}\ (B^+\to K^+f_0(980);\ f_0(980)\to \pi^+\pi^-)\ = \
(9.6^{+2.5+1.5+3.4}_{-2.3-1.5-0.8})\times 10^{-6}. \label{f0krate}
\end{equation}
The preliminary result by BaBar \cite{babar1} reads
\begin{equation}
{\rm Br}\ (B^+\to K^+f_0(980);\ f_0(980)\to \pi^+\pi^-)\ = \
(9.2\pm1.2^{+2.1}_{-2.6})\times 10^{-6}. \label{f0krate1}
\end{equation}
This large fraction of $f_0(980)$ (3 times larger than $\rho^0$) 
is a first hint for the gluonic
affinity of this meson as well.

\begin{table}[t]
\caption{Dominant contributions for 
$B$ decays into scalar (S) + pseudoscalar (P) or
vector (V) particles: penguin amplitudes $p_{AB}$ (normalized to 1
in each sector), exchange  and
gluonic amplitudes  $\beta\equiv\beta_{PS},\beta'\equiv\beta_{VS}$ 
and $\gamma_P,\gamma_S$ 
resp.; also approximate forms for
mixing angles $\varphi_S=\varphi_P$ and $\beta=-\beta'=1$; notations
$f_0\equiv f_0(980),\ f_0'\equiv f_0(1500)$ and $K^*_{sc}\equiv K^*_0(1430)$.}
$
\begin{array}{llcllc}
\hline
B^0\to & B^+\to & \text{normalization to}&
  B^0\to & B^+\to & \text{normalization to}\\
P+S & P+S &  p_{PS} & V+S & V+S &  p_{VS} \\
 \hline
K^+a^- & K^0 a^+ & 1 & K^{*+}a^- & K^{*0} a^+ & 1 
\\
K^0a^0 & K^+ a^0 & \frac{1}{\sqrt{2}} & K^{*0}a^0 & K^{*+} a^0 &
   \frac{1}{\sqrt{2}} 
\\
K^0f_0 & K^+f_0 & \frac{1}{\sqrt{2}}(1+2\gamma_S)\sin \varphi_S &
  K^{*0}f_0 & K^{*+}f_0 & \frac{1}{\sqrt{2}}(1+2\gamma_S)\sin \varphi_S\\
       &        & \quad +(\beta + \gamma_S)\cos \varphi_S &
       &        & \quad  +(\beta' + \gamma_S)\cos \varphi_S  \\
     & & \approx  \frac{1}{\sqrt{6}}(3+4\gamma_S) & 
     & & \approx  \frac{1}{\sqrt{6}}(-1+4\gamma_S)\\
K^0f_0' & K^+f_0' & \frac{1}{\sqrt{2}}(1+2\gamma_S)\cos \varphi_S & 
     K^{*0}f_0' & K^{*+}f_0' & \frac{1}{\sqrt{2}} (1+2\gamma_S)\cos
      \varphi_S\\
      & & \quad   -(\beta + \gamma_S)\sin \varphi_S &
      & & \quad  -(\beta' + \gamma_S)\sin \varphi_S  \\
     & & \approx  \frac{1}{\sqrt{3}} \gamma_S &
         & & \approx  \frac{1}{\sqrt{3}} (2+\gamma_S)\\
\pi^-K^{*+}_{sc} & \pi^+K^{*0}_{sc} & \beta & 
    \rho^- K^{*+}_{sc} &  \rho^+ K^{*0}_{sc} & \beta
\\
   \pi^0K^{*0}_{sc} & \pi^0K^{*+}_{sc} & \frac{1}{\sqrt{2}} \beta &
   \rho^0K^{*0}_{sc} & \rho^0K^{*+}_{sc} & \frac{1}{\sqrt{2}} \beta
\\
\eta K^{*0}_{sc} & \eta K^{*+}_{sc} &
\frac{1}{\sqrt{3}}(-1+\beta+\gamma_P)&
     \omega K^{*0}_{sc} & \omega K^{*+}_{sc} & \frac{1}{\sqrt{2}} \beta 
\\
\eta' K^{*0}_{sc} & \eta K^{*+}_{sc} &
    \frac{1}{\sqrt{6}}(2+\beta+4\gamma_P)&
   \phi  K^{*0}_{sc} & \phi K^{*+}_{sc} & 1\\
\hline
  \end{array}
$
\label{tab:scalars}
\end{table}

It is clear that a more definitive answer requires an analysis 
similar to the one with $\eta'$ for scalar (S) particles
as well. To this end we have written down in Table
\ref{tab:scalars} the amplitudes for
the decays $B\to PS$ and $B\to VS$ in the approximation as above,
keeping only
the QCD penguin and gluonic amplitudes $p_{PS},p_{VS}$ and $\gamma_Sp_{PS},
\ \gamma_Sp_{VS}$ $(\gamma_S\equiv\gamma_{PS}\equiv \gamma_{VS})$. We assume the scalar nonet 
with states as in
Ref. \cite{mo}. Because of the large phase space in $B$ decays there is a
good chance to identify the scalar particles belonging to the nonet of
lowest mass with little background from crossed decay channels. 

In order to determine the gluonic production amplitude $\gamma_S$ for 
 $f_0(980)$ one
needs to identify additional channels, in particular with 
$a(980)$ and $K^*(1430)$,
the suggested partners of pion and kaon, respectively. 
The latter state has apparently been observed by
BELLE \cite{belle} (called $K_X(1400)$) albeit with large error
(Br($B^+\to K_{sc}^{*0}\pi^+)\sim (21.7^{+7.6}_{-11.3})\times 10^{-6}$). 
Measurements together with
$K^*(890)$ can reveal the different effects the gluonic amplitude has
on $f_0(980)$ and $f_0(1500)$, the suggested partners of $\eta'$ and $\eta$. 
Given several rates for pure quark states 
the scalar mixing angle
$\varphi_S$ defined through
\begin{equation}
f_0(980)=n\overline n \sin \varphi_S + s\overline s \cos \varphi_S,\quad
f_0(1500)=n\overline n \cos \varphi_S - s\overline s \sin \varphi_S
\label{mixings}
\end{equation}
with $n\overline n=(u\overline u +d\overline d)/\sqrt{2}$
can be determined as well. Our choice \cite{mo} is
$\sin\varphi_S=1/\sqrt{3}$ as in the pseudoscalar nonet, i.e.
$\varphi_P\approx\varphi_S$. 

The strong appearence of $f_0(980)$ in the final state with intermediate
gluons points toward a flavour singlet component. 
As further test we note that
in the approximation of Table \ref{tab:scalars}
the decay $B\to Kf_0(1500)$ should be suppressed in analogy to $K\eta$.
Indeed, there is no signal from $\pi^+\pi^-$ or $K^+K^-$ at this mass
in the BELLE data \cite{belle} (assuming $f_0(1500)$ is the partner in the 
same nonet), however, there is a
large branching ratio of $f_0(1500)$ also into $4\pi$. 
On the other hand,
$f_0(1500)$ should show up together with $K^*$. Furthermore we note that
$f_0(980)$ could interfere destructively with the background which would
result in a much larger decay rate. This occurs for the amplitude 
$T_B+T_{f_0}e^{2i\phi_B}$ in case of background phase $\phi_B\sim \pi/2$
($|T_B|\ll|T_{f_0}|$ here, $T=|T|e^{i\phi}$) as we expect for our glueball interpretation below. 


Next we consider a possible mechanism
for the $B\to f_0(980) K$ decay similar to  $B\to \eta'K$ assuming a direct
coupling to a two gluon state. 
Then we expect that
the gluonic  couplings of $\eta'$ and $f_0$ are proportional 
to the corresponding processes with photons.
In consequence, the ratios
\begin{equation}
R_1=\frac{{\rm Br}(B\to f_0(980)K)|_{\text{gluonic}}}
    {{\rm Br}(B\to
\eta'K)|_{\text{gluonic}}}=\frac{|\gamma_Sp_{PS}|^2}{|\gamma_Pp_{PP}|^2}, 
\quad 
R_2=\frac{\Gamma(f_0(980)\to \gamma\gamma)}{\Gamma(\eta'\to \gamma\gamma)}
\label{r1r2}
\end{equation}    
should be of comparable size if indeed the  mixing angles
$\varphi_P\approx\varphi_S$, then the quark charge factors
cancel in $R_2$. 
Taking $\Gamma(\eta'\to \gamma\gamma)=(4.29\pm 0.15)$ keV
 and $\Gamma(f_0(980)\to \gamma\gamma)=(0.39^{+0.10}_{-0.13})$ keV 
\cite{pdg} we obtain $R_2\sim 0.09  \pm 0.03$.

With the assumption $R_1\approx R_2$ and together with the decay rate
 (\ref{f0krate}) or (\ref{f0krate1}) using Table \ref{tab:pseudo} for
$\eta'K$
we can actually determine $\gamma_S$ and $p_{PS}$ taken as real parameters.
With  Br$(B^+\to f_0(980)K^+)\sim (14\pm 4)\times 10^{-6} $
 after correction for
$\pi^0\pi^0$ decay of $f_0$ and neglecting $K\overline K$ we find two
solutions
\begin{equation}
\text{A:}\quad \gamma_S=-0.17,\ p_{PS}^2=15\times
10^{-6};\qquad 
\text{B:}\quad \gamma_S=0.3,\ p_{PS}^2=5\times
10^{-6}
\label{solscal}
\end{equation}
According to Table \ref{tab:scalars} the $K^{*0}_{sc}\pi^\pm$ and $Ka^\pm$ 
rates are of order $p_{PS}^2$. If we take the quoted BELLE result 
on $K^{*0}_{sc}\pi^+$ into account, then  Solution A is favoured. In this
solution  the rates for 
$K^*f_0$, $K^*f_0'$ are  $\sim 7\alpha^2_S\times 10^{-6}$ 
and $\sim 17\alpha^2_S\times 10^{-6}$ resp. whereas the same rates in Solution B
are  $\sim 0.0\times 10^{-6}$ and $\sim 9\alpha^2_S\times 10^{-6}$ where
$\alpha_S=p_{VS}/p_{PS}$.

{\it Total rate for gluonic decays}\\ 
Next we compare the rates for $f_0$ and $\eta'$ production with the
total rate $b\to sg$ in (\ref{btosg}).
 CLEO \cite{cleo1} has measured the inclusive non-charm decay
Br$(B\to \eta'+X)\ = \ (6.2^{+2.1}_{-2.6})\times 10^{-4}$,
where the signal refers to the region $2.0<p_{\eta'}< 2.7$ GeV
of the $\eta'$ momentum. Identifying the non-charm rate with $X_s$ according
to the SM and adding the exclusive $\eta'K$ rate we obtain
the inclusive rate 
${\rm Br}(B\to \eta'+X_s)\ \sim \ 7.0\times 10^{-4}$,
so the total inclusive $\eta'X_s$ rate is about 9 times larger than the
exclusive $\eta'K$ rate.
We take the gluonic part as in (\ref{etaprgluon}) and add a
gluonic contribution for $f_0(980)$ of fraction $R_2\sim 10\%$. 
 Then we find for the fully inclusive contribution of these  decays 
\begin{equation}
{\rm Br} (B\to \eta',f_0(980))|_{\rm gluonic})\sim (1.5\ldots 3.5)\times
   10^{-4}.
\label{glumeson}
\end{equation}  
Hence, these decays cannot contribute more than $\sim (3-7)\% $
of the  expected $b\to s g $ rate of $5\times 10^{-3}$ 
\cite{greub}.
We will argue below that glueball production does provide the dominant
part of the $b\to sg$ decay with ``real'' gluon.

\section{Gluon jet fragmenting into $f_0(980)$}
Gluonic mesons should also be found as leading particles in gluon jets
\cite{pw,fritzsch,gl,mo1,sz}.
Following the proposal in ref. 
\cite{mo1} the 3-jet events obtained by DELPHI
at LEP have been used to isolate the
leading component in gluon jets.
Events have been selected with a rapidity gap in this jet 
and first results have been
presented \cite{bm}. The charge of the
leading component of the gluon jet beyond the rapidity gap
has been compared with the MC simulation.
Whereas the quark jets showed good agreement with this calculation the gluon
jets had an excess of jets with charge $Q=0$ as expected for an extra
gluonic component. 
The $\pi^+\pi^-$ mass spectrum showed the $Q=0$ excess spread over a
large mass range with considerable fluctuations, but in the low mass region
a significant peak is found for $f_0(980)$, 
absent in quark jets. This is a strong hint at the gluonic affinity of
$f_0(980)$, i.e. its flavour singlet nature. 

Results are desirable from energetic jets with large gaps and good separation 
from neighbour jets to minimize background. Such jets 
 with high $p_T$ are 
produced also in hadronic or $ep$ 
collisions.
 

\section{Glueball production in $B$ decays}
Besides the observation of the strong $f_0(980)$ signal in charmless $B$
decays, there is another interesting feature in the decays $B^+\to
K^+\pi^+\pi^-$
and $B^+\to K^+ K^- K^+$ observed by the BELLE collaboration \cite{belle}. 
The latter channel shows a broad enhancement in the $ K^+ K^-$ mass
spectrum in the region $1.0-1.7$~GeV. The flat 
distribution in the Dalitz plot of
these events suggests this object to be produced with spin $J=0$. 
Its contribution 
 has been parametrized as scalar state $f_X(1500)$ with mass $M=1500$ MeV
and $\Gamma=700$ MeV. There is no
sign of a particular narrow resonance such as $f_0(1500)$ with width of
about 100 MeV; this latter 
state should be seen more clearly in the $\pi\pi$ spectrum,
given the small ratio $\Gamma(K\overline
K)/\Gamma(\pi\pi)=0.19\pm0.07$ \cite{cbc} but there is no sign here either
in the same experiment.

In the $\pi\pi$ mass spectrum in 
$B^+\to K^+\pi^+\pi^-$ there are enhancements around
$f_0(980)$ in the region 0.7-1.4 GeV. In this case, because of high
background and smaller statistics, the spin is not so obvious but at least
some $J=0$ component is apparently present. The enhancement above 1 GeV has
been related to $f_0(1370)$. Similar results have been reported by 
BaBar \cite{babar1}.

In our earlier study of low mass scalars \cite{mo} we interpreted the
$\pi\pi$ S wave as being dominated by a very broad scalar state which
interferes destructively with  narrow $f_0(980)$ and $f_0(1500)$ (``red
dragon'') and extends in mass up to about 1600 MeV. This broad object,
corresponding to $f_0(400-1200)$ and $f_0(1370)$ listed by the PDG, 
we classified as scalar glueball $gb(1000)$ after the other low mass scalar
states have been filled into the $q\overline q$ nonet. We argued that the
Breit Wigner phase motion for $f_0(1370)$ has not been demonstrated clearly
enough to require an extra state.

We consider the enhancements in $\pi\pi$ and $K\overline K$ 
observed by BELLE as a new, very
clear manifestation of this broad scalar glueball. Whereas the center of the
peak in $\pi\pi$ is closer to 1 GeV, it is shifted to higher mass in the
$K\overline K$ channel. 
Because of the
large width the branching ratios into different channels 
vary strongly with mass depending on the respective thresholds. The shape of the
mass spectrum is also expected to vary from one reaction to another because
different kinematic and dynamic factors may apply.

In order to relate different channels and to obtain an estimate of the total
glueball production rate we consider the following decay scheme. The
glueball decays first into $q\overline q$ pairs (possibly glueballs)
\begin{equation}
gb\to u\overline u + d\overline d + s\overline s\quad (+gb\ gb) 
\label{gbdecay}
\end{equation}
subsequently, each of these $q\overline q$ 
pairs recombines with a newly created pair
$u\overline u$, $d\overline d$ or $s\overline s$ where $s\overline s$ is
produced with amplitude $S$ ($|S|<1$). In this way the 2-body channels 
$gb\to q\overline q'+\overline q q'$ are
opened, at low energies just pairs of pseudoscalars. They are produced
with probabilities
\begin{equation}
\vspace{-0.8cm}
\begin{array}{cccccc} 
\pi^+\pi^- & \pi^0\pi^0 & K^+ K^- & {K^0 \overline K}^0 & \eta\eta &
     \eta'\eta'\\
 2 & 1 & 2 & 2 & 1 & 1 \\
 2& 1 & \frac{1}{2}|1+S|^2 & \frac{1}{2}|1+S|^2 & \frac{1}{9}|2+S|^2 &
  \frac{1}{9}|1+2S|^2\\
\label{gb2body}
\end{array}
\end{equation}

\vspace{-0.6cm}
\noindent The first row corresponds to $U(3)$ symmetry ($S=1$), the second
row to arbitrary $S$; $\eta,\eta'$ mixing is assumed as
above. With increasing glueball mass 
the $q\overline q$ pairs can
decay also into pairs of vector mesons or of other states but the total
rates in (\ref{gb2body}) are assumed to remain unaltered.

We study first the mass region 1.0-1.7 GeV.
In this region the pseudoscalars alone saturate the $K\overline K$ 
rate in (\ref{gb2body})
as $K^*\overline K$ is forbidden by parity and $K^*{\overline K}^*$ is 
kinematically suppressed.
Another possible decay is $\eta\eta$, contributions from higher mass
isoscalars ($\omega\omega$) are only possible at the upper edge of the
considered mass interval. The decay  $\eta'\eta'$ is kinematically
forbidden.
On the other hand, the $\pi\pi$ channel
can get contributions from higher states, in particular $\rho\rho$
which becomes effective in the mass region above 1300 MeV. 

Next we estimate the total glueball rate. 
In Table \ref{tab:rates} we start from the observed $K\overline K$ rate 
in the mass interval 1.0-1.7 GeV and
derive using (\ref{gb2body}) the rates for $\eta\eta$ and ``$\pi\pi$''
where the latter includes $\rho\rho$. This yields the rate ($124\pm37)\times
10^{-6}$.
We may compare the prediction for
``$\pi\pi$'' with the observed $\pi\pi$ rate in the region around
$f_0(1370)$ which we consider as part of the glueball.
Its decay properties are not well known,
but there is a considerable fraction into $4\pi$, in particular 
$\rho\rho$. If the result from
the  extrapolation of $\pi\pi$ alone  
($\sim 64\times 10^{-6}$) is  multiplied by factor 2 in order to account for
the larger mass interval of $K\overline K$, then both results 
(predicted and extrapolated) are consistent
within the very large errors (Table \ref{tab:rates}).

\begin{table*}
\caption{Observed \protect\cite{belle} 
and corrected/expected branching ratios for 
$B\to K^+gb(0^{++})$ decays for
different 2-body channels (with $S=0.8$).
}
\vspace*{0.1cm}
$
\begin{array}{clll}
 \hline
 \mbox{mass range}& \mbox{Br}_{B^+\to RK^+}\times &
       \mbox{corr./exp. Br} & \mbox{comments}\\
 \mbox{[GeV]} & \times \mbox{Br}_{R\to h^+h^-} [10^{-6}] & [10^{-6}] & \\
\hline
1.0-1.7 
   & K^+K^-: 27.6\pm 4.9   & K\overline K \quad 55.2\pm 9.8
  &  \mbox{factor}\  2 \ (\text{isospin})
 \\ 
& & \eta\eta \qquad 17.3\quad  & K\overline K/\eta\eta=3.2
\quad(\text{Eq.}\protect\ref{gb2body})
  \\
& & ``\pi\pi\text{''} \quad 51.8  &  K\overline K/\pi\pi=1.08  
  \ (\text{Eq}.\protect\ref{gb2body}) 
\\
\hline
 & & \mbox{all}\qquad 124\pm 37 & 
\\
\hline\hline
1.0-1.3  
   & \pi^+\pi^-: 11.1^{+8.0}_{-4.5} & \pi\pi \qquad
16.7^{+12.0}_{6.7} & \mbox{factor}\ 3/2 \ (\text{isospin})
\\
\hline
 & & \mbox{all:}\quad \sim 64^{+65}_{-36} & \pi\pi/\mbox{all} =0.26\pm0.09
\ \protect\cite{bugg} 
\\
\hline \hline
0.7-1.0 
       & & \sim 8 & \mbox{estimate}\\
\hline\hline
0.7-1.7 
     & & 132 & B\to K^+gb(0^{++})\ \mbox{total}\\
\hline
  \end{array}
$
\label{tab:rates}
\end{table*}

Finally, from the total rate $B^+\to K^+gb(0^{++}) \sim 132\times 10^{-6}$ 
in Table \ref{tab:rates} obtained by adding the low mass interval 
we may estimate the total inclusive rate by applying the same
factor 9 as in case of $\eta'$ in the extrapolation $K^+\to X_s$ and find
\begin{equation}
\mbox{Br} ( B^+\to gb(0^{++})+X_s) \sim 1.2\times 10^{-3} \label{glutot}
\end{equation}
If we add the gluonic $\eta'X_s$ and  $f_0(980)X_s$ contributions 
in (\ref{glumeson}) then we estimate the total production
of observed gluonic mesons as
\begin{equation}
\mbox{Br} (B^+\to gb(0^{++})+f_0+\eta'+X_s) \sim (1.5\pm0.5)\times 10^{-3} 
\label{glumestot}
\end{equation}
which is of the same size as the leading order result for the process
$b\to sg$ in (\ref{btosg}) and about 1/3 of the full rate obtained in NLO.

Besides the scalar glueball other glueballs should be produced as well.
For orientation, it is plausible to assume 
\begin{equation}
\mbox{Br}(B\to gb(0^{++})+X_s)\quad \approx\quad 
   \mbox{Br}(B\to gb(0^{-+})+X_s)
\end{equation}  
This is obtained if one assumes a symmetry under chromoelectric-magnetic rotation
$F\to \tilde{F}$ for the operators $F^2$ and $F\tilde{F}$ and 
approximately neglects the breaking of this symmetry.
In this case the scalar and pseudoscalar gluonic mesons would add up to
about $B\to (J=0)\ \mbox{glue mesons} \sim 3\times 10^{-3}$ which is close to the
total gluonic decay rate $b\to sg$. Given the errors in these
estimates the counted decays could actually saturate the total rate,
alternatively, there is room for glueballs with higher spin or hybrid states.

\section{Conclusions}

The large decay rate $B\to f_0(980) K$ and the excess of $f_0(980)$
in the leading part of the gluon jet suggest the gluonic affinity of this
meson similar to $\eta'$ 
and therefore its flavour singlet nature. The further study of  $B$ decays into
scalar particles could be of invaluable help in establishing the members
of the still controversial $0^{++}$ multiplet and its mixing. 
We also remark that the large exclusive decay rate for $f_0(980)$ 
makes a 4-quark or molecular hypothesis unplausible as formfactors
are expected rather to suppress this decay at the high energy of the $B$ meson. 
Similarly, measurements of $f_0(980)$ (also $\eta'$) 
in gluon jets with larger rapidity 
gaps for background suppression are desirable, possibly also from $pp$ and
$ep$ collisions.

Our interpretation of $B\to K\overline KK$ decays 
in terms of glueball production
should be tested by observing the predicted missing channels 
($\eta\eta$ and $4\pi$)
with the rates expected for the flavour singlet (Table \ref{tab:rates}).
 
The flavour singlet nature of $f_0(980)$ does not necessarily
imply a large mixing with a scalar glueball, such a large mixing is absent
also in case of $\eta'$. The near singlet flavour mixing of $\eta'$ and
$f_0(980)$ is intrinsically due to their gluonic couplings. In fact,
taking the results of our glueball analysis for granted with rate
$132\times 10^{-6}$ then an upper limit for mixing results by
assuming the full rate
of $f_0(980)$ production ($14\times 10^{-6}$) to be 
due to mixing with glueball;
then the mixing angle would be $\sin^2 \varphi_g\lsim (14/132)$ or
$\varphi_g\lsim 20^\circ$.

Ultimately there is the question of how the $b\to sg$ decay is realized
by hadronic final states. The large rate for the  $0^{++}$ glueball we
obtain
suggests the intriguing possibility 
that it could be saturated by gluonic mesons. In the next 
step it will be interesting to search for the $0^{-+}$ glueball which
could decay into $\eta\pi\pi$ and $K\overline K\pi$. The candidate 
of lowest mass would be $\eta(1440)$ which is strongly produced
in radiative $J/\psi$ decays.

\end{document}